\documentclass{article}
\usepackage{a4wide}
\usepackage[utf8]{inputenc}

\usepackage{graphicx}
\usepackage{caption}
\usepackage{subcaption}
\usepackage[colorlinks=true,linkcolor=black,anchorcolor=black,citecolor=black,filecolor=black,menucolor=black,runcolor=black,urlcolor=black]{hyperref}

\usepackage{abstract}

\usepackage{authblk}

\usepackage{gensymb}

\usepackage[citestyle=ieee,firstinits=true,backend=biber]{biblatex}
\renewbibmacro{in:}{}

\title{Hybrid integrated near UV lasers using the deep-UV $\mathrm{Al_{2}O_{3}}$ platform}

\author[1]{C.A.A. Franken}
\author[2]{W.A.P.M. Hendriks}
\author[1,3]{L.V. Winkler}
\author[2]{M. Dijkstra}
\author[4]{A.R. do Nascimento Jr.}
\author[1]{A. van Rees}
\author[2]{M.R.S. Mardani}
\author[5]{R. Dekker}
\author[4]{J. van Kerkhof}
\author[1]{P.J.M. van der Slot}
\author[2]{S.M. García-Blanco}
\author[1]{K.-J. Boller}
\affil[1]{Laser Physics and Nonlinear Optics Group, Department of Science and Technology, MESA\textsuperscript{+} Institute of Nanotechnology, University of Twente, Enschede, The Netherlands}
\affil[2]{Integrated Optical Systems, Department of Science and Technology, MESA\textsuperscript{+} Institute of Nanotechnology, University of Twente, Enschede, The Netherlands}
\affil[3]{TOPTICA Photonics, Gr{\"a}felfing, Germany}
\affil[4]{PHIX B.V., Enschede, The Netherlands}
\affil[5]{LioniX International B.V., Enschede, The Netherlands}

\date{February 22, 2023}

\begin{document}
\maketitle

\vspace{-0.25cm}
\begin{abstract}%
\noindent \textbf{Hybrid integrated diode lasers have so far been realized using silicon, polymer, and silicon nitride ($\mathbf{Si_{3}N_{4}}$) waveguide platforms for extending on-chip tunable light engines from the infrared throughout the visible range. Here we demonstrate the first hybrid integrated laser using the aluminum oxide ($\mathbf{Al_{2}O_{3}}$) deep-UV capable waveguide platform. By permanently coupling low-loss $\mathbf{Al_{2}O_{3}}$ frequency-tunable Vernier feedback circuits with GaN double-pass amplifiers in a hermetically sealed housing, we demonstrate the first extended cavity diode laser (ECDL) in the near UV. The laser shows a maximum fiber-coupled output power of 0.74 mW, corresponding to about 3.5 mW on chip, and tunes more than 4.4 nm in wavelength from 408.1 nm to 403.7 nm. Integrating stable, single-mode and tunable lasers into a deep-UV platform opens a new path for chip-integrated photonic applications.}
\end{abstract}
\vspace{0.25cm}
\noindent A wide range of emerging photonic applications in the ultraviolet requires chip-sized integrated laser sources in the ultraviolet with wide tunability and high coherence. Specifically, such lasers would unlock increased integration density and upscaling in integrated quantum photonics \cite{Moody2022,Mehta2020,Niffenegger2020}, UV spectroscopy \cite{Zybin2005,Liu2022}, UV biophotonics \cite{Lin2022}, and multiple UV optical clock transitions \cite{Ludlow2015}.\par
Heterogeneous and hybrid integrated lasers provide a wide tuning range and high coherence at longer wavelengths, while maintaining small size and high efficiency. In these devices, photonic integrated circuits (PICs) are used to spectrally filter and couple back light from III-V semiconductor optical amplifiers to impose tunability and improve coherence. In the infrared telecom range, such lasers are now providing tunability across the entire gain bandwidth, and allow for coherence levels which are becoming comparable to advanced bulk laser systems. This evolution in performance became possible by introducing a change of paragdigm in chosing the optical materials for PIC fabrication. The initially used semiconductor feedback circuits, intended to address the infrared telecom  range, offered only small material bandgaps, such as in Si waveguides (1.1 eV) or InP waveguides (1.3 eV). The situation changed dramatically when low-loss waveguides with a much wider bandgap became available. An example of such a waveguide platform is silicon nitride ($\mathrm{Si_{3}N_{4}}$) embedded in silicon oxide ($\mathrm{SiO_{2}}$) \cite{Roeloffzen2018,Blumenthal2018} which provides a bandgap as high as 3.3 eV \cite{Kruckel2017}. This reduced both linear and nonlinear material losses, enabled highly frequency selective waveguide circuits and allowed to maximize laser coherence with extended on-chip photon lifetimes. Recently, the wide bandgap also enabled the first realization of hybrid integrated lasers in the visible range with mW-level fiber-coupled output \cite{Franken2021,Winkler2023}. Employing separate feedback chips for spectral narrowing and frequency pulling with Fabry-Perot lasers resulted in visible tunable output as well. Milliwatt level fiber-coupled powers were  obtained in the blue range (450-460 nm), about 500 µW in the violet \cite{Wunderer2023}, however, at the very end of the visible the power was restricted to the 100 and 10-µW regimes \cite{Siddharth2022,Corato-Zanarella2022}, respectively.\par
For reaching out to shorter wavelengths in the UV range, it has recently been discussed whether employing silicon nitride would still be appropriate \cite{Tran2021a}. However, a straightforward extension to the UV range using $\mathrm{Si_{3}N_{4}}$ feedback circuits, especially combined with 5-eV GaN diodes at 250 nm can be excluded \cite{Liu2018}. Moving into these ranges would increase the waveguide propagation losses strongly, as the photon energy approaches, matches and supersedes the 3.3-eV silicon nitride bandgap (corresponding to 380 nm) \cite{Agaskar2019,Morin2021}. Ultimately, if losses become too high, the feedback circuit loses its filter and feedback function, and control over tuning and coherence fails. Even below the bandgap energy, $\mathrm{Si_{3}N_{4}}$ has shown to be susceptible to nonlinear excitation \cite{Porcel2017,Billat2017a}, while UV irradiation creates defects that introduce absorption also at below-bandgap wavelengths \cite{Neutens2018}.\par
Clearly, these considerations and observations require to introduce a deep-UV capable platform to extend the complex functionalities of laser feedback circuits into the UV, such that the material bandgap remains much wider than the targeted UV wavelength. The most promising materials for this task are aluminum nitride (AlN) and aluminum oxide ($\mathrm{Al_{2}O_{3}}$) because of their huge bandgaps of 6 eV (200 nm) and 7.6 eV (165 nm) respectively, \cite{West2019c,Lu2018}. As the $\mathrm{Al_{2}O_{3}}$ platform does not suffer from anisotropy and high sidewall roughness due to AlN crystallinity and provides a wider spectral coverage it seems more suitable for deep-UV applications. Further indicators of suitability as a laser feedback platform are that simple low-loss waveguide components have already been fabricated \cite{Hendriks2021b,West2019c,Lin2022}, extending the established range of fabrication tools for $\mathrm{Al_{2}O_{3}}$-based infrared waveguide lasers \cite{Hendriks2021a,Bonneville2020}. Tight guiding in $\mathrm{Al_{2}O_{3}}$ high-Q ridge waveguide resonators has been demonstrated so far only in the near-UV, showing intrinsic qualitity factors of Q $>$ 470.000 \cite{West2019c}. Comparably low straight propagation losses were found in the ultraviolet as well (3 dB/cm at 360 nm \cite{Lin2022}).\par
These features underline the potential of $\mathrm{Al_{2}O_{3}}$ waveguides for spectral control of chip-integrated UV lasers, however, in this wavelength range there are additional challenges. Due to the high photon energy, losses and damage can occur at all UV-exposed surfaces and interfaces such as the facets of semiconductor optical amplifiers, i.e., GaN. The reasons are photo-induced chemical processes with atmospheric moisture and gases that form surface
defects or absorptive layers \cite{Uddin2005,Schoedl2005a}. This degradation needs to be counteracted, such as by hermetic sealing for shielding from ambient humidity. In the infrared, hybrid integration with transparent epoxy for permanent chip-to-chip bonding, heterogeneous integration, and packaging solely aim on long-term frequency stability, robustness and electro-optical connectivity. With UV lasers, however, chip-to-chip bonding, chip-to-fiber bonding, and photonic packaging \cite{Ranno2022} has to be  UV compatible as well. Similarly, photonic wirebonding \cite{Xue2021} and heterogeneous integration via direct bonding \cite{Liang2010} or transfer printing \cite{Wang2017} would have to involve UV-suitable materials, such as AlN as buffer layers \cite{Sun2016b}. Otherwise, stable long-term laser operation can not be achieved.\par
\begin{figure}[!t]
    \centering
    \makebox[\textwidth][c]{\includegraphics[width=180mm]{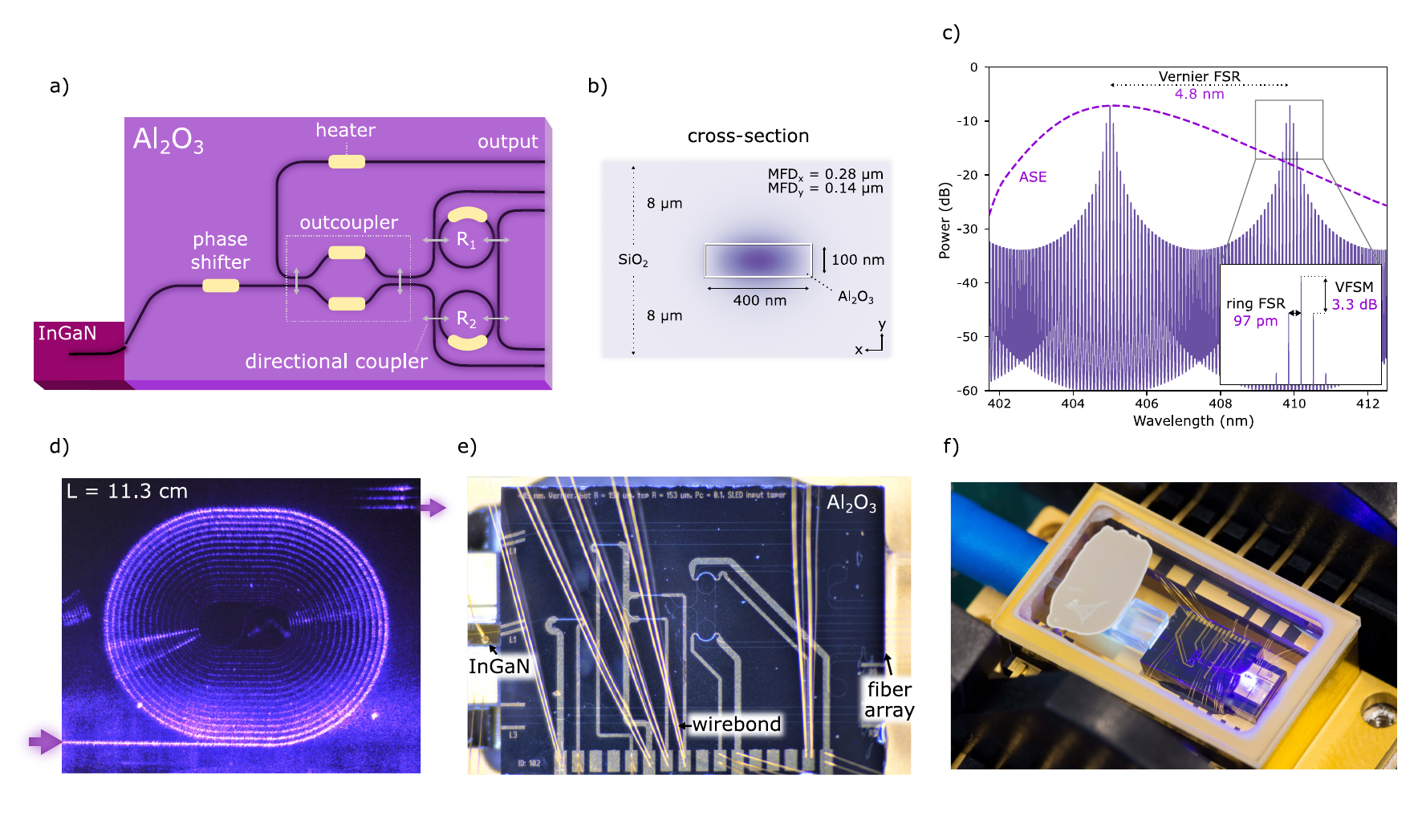}}%
    \caption{\textit{a) Schematic showing the functional design of the laser, with the InGaN amplifier edgecoupled to the $Al_{2}O_{3}$ feedback chip. b) Cross-section of the $Al_{2}O_{3}$ waveguide embedded in $SiO_{2}$ showing the calculated intensity profile of the optical mode (at 405 nm), with a horizontal and vertical modefield diameter (MFD) of 0.28 $\mu m$ and 0.14 $\mu m$ respectively. c) Transmission function (solid line) for light resonating with both rings in the Vernier filter. A smoothed curve for the measured amplified spontanous emission (ASE) of the amplifier driven at 55 mA (dashed line) shows the Vernier filter can select a single wavelength in the amplifier emission bandwidth. d) Light at 405 nm injected in an 11.3-cm spiral structure. A series of spirals and paths of different length were used to measure the straight propagation loss of $\mathrm{\alpha}$ = 2.8 $\pm$ 0.3 dB/cm. e) Top-down microscopic picture of the hybrid laser, showing the waveguides, heaters, wirebonds and location of the InGaN amplifier and fiber array in detail. f) The hybrid integrated and packaged laser in a hermetically sealed nitrogen environment. Here, the laser is driven with a pump current above threshold. The white substance on the bottom of the glass lid is a getter material which absorbs any remaining volatile gasses and moisture in the package.}}
    \label{fig:1}
\end{figure}

Here we present a chip-scale hybrid integrated laser, where for the first time the deep-UV capable $\mathrm{Al_{2}O_{3}}$ platform is employed (see in Fig. \ref{fig:1})\cite{Franken2023}. Using CMOS compatible technology, $\mathrm{Al_{2}O_{3}}$ waveguides embedded in a $\mathrm{SiO_{2}}$ cladding are fabricated for tightly guided single-mode propagation, minimized side-wall scattering, and minimized bend radiation loss. For obtaining single wavelength laser oscillation, two sequentially coupled $\mathrm{Al_{2}O_{3}}$ microring resonators are fabricated and connected in Vernier configuration \cite{Coldren2012}, which provides frequency selective feedback to a near UV InGaN double-pass semiconductor amplifier. Thin-film micro-heaters enable thermo-optical laser tuning and provide tunable output coupling. Hybrid integration serves for permanent optical coupling of amplifier, feedback chip and output fibers. The packaged chip assembly is based on hermetic sealing to yield protection against degradation for long-term operation. The laser generates a fiber-coupled output power of 0.74 mW, which corresponds to an on-chip power of 3.5 mW, and is tunable in wavelength more than 4.4 nm from 408.1 nm to 403.7 nm.
\subsection*{Waveguide cross-section and fabrication}
We chose to make use of $\mathrm{Al_{2}O_{3}}$ cores embedded in a $\mathrm{SiO_{2}}$ cladding to provide a maximum electronic bandgap (7.3 eV \cite{West2019c} and 9.3 eV \cite{Weinberg1979}, respectively). To identify a suitable core cross-section for fabrication, preceding infrared and UV measurements were consulted and used for scaling. With 400 nm thick and 2000 nm wide straight $\mathrm{Al_{2}O_{3}}$ waveguides we measured losses of 0.05 dB/cm at a wavelength of 1550 nm. In the UV at 377 nm, for a 170 nm thick $\mathrm{Al_{2}O_{3}}$ film deposited with a reactive sputtering process and followed by chemical mechanical polishing, we measured slab propagation losses of 0.6 $\pm$ 0.3 dB/cm \cite{Mardani2022}. Choosing a conservative approach for this first-time $\mathrm{Al_{2}O_{3}}$-based hybrid integrated diode laser, we decided for a near UV target wavelength of 405 nm. In that range, suitable double pass gallium nitride amplifiers, as well as equipment for laser output characterization, was available. Scaling the measured UV and IR losses to the target wavelength range predicts losses between 1 and 4 dB/cm for 400 nm thick waveguides. Selecting a thinner waveguide would reduce propagation loss via reduced sidewall scattering while maintaining tight guiding for UV wavelengths.\par
Choosing a proper waveguide cross-section is of central importance to design low-loss and fabrication tolerant waveguide components. This applies to curved waveguides in small ring resonators for spectral filtering with wide free spectral range, where the cross-section and ring radius determines the bending loss and thus the filter $Q$-factor. Also the design of directional couplers depends on the cross-section, as the coupling strength is influenced by the coupler length and by the fabrication tolerance of the gap between the waveguide cores. The cross-section to be selected also needs to restrict propagation to a single transverse mode with a polarization matching that of the amplifier ($\mathrm{TE_{00}}$), while efficient coupling to the diode amplifier and output fibers should be enabled by modematching through inverse tapering at the selected waveguide thickness. Using measured losses, scaling and numerical simulations (see Methods) a waveguide cross section of a 400 nm wide by 100 nm thick $\mathrm{Al_{2}O_{3}}$ core embedded in a $\mathrm{SiO_{2}}$ cladding is chosen (see Fig. \ref{fig:1}b).\par
The fabrication employs common CMOS compatible techniques on a wafer scale (see Methods for details). With a mask design for a 10-cm wafer, 127 chips are fabricated carrying an extensive range of test structures from waveguide spirals (Fig. \ref{fig:1}d) for loss measurements to Mach-Zehnder interferometers (MZI) to test the heater functionality. Various Vernier feedback circuits as shown in Fig. \ref{fig:1}a are fabricated with variation in ring radii, tapers and coupling coefficients between ring resonators and bus waveguides. Characterizing the individual test structures and considering the measured loss at the nominal wavelength allows us to fabricate and select suitable feedback circuits for hybrid integration with amplifiers, while requiring only a single wafer run.
\subsection*{Design of $\mathbf{Al_{2}O_{3}}$ feedback chip}
Employing $\mathrm{Al_{2}O_{3}}$ waveguides for novel near UV laser feedback circuits requires to adapt the design of crucial circuit functions. One function is providing tunable optical filtering to the amplifier with the goal to impose wavelength tunable laser output with high side-mode suppression. Tuning reliably across the entire gain bandwidth is easily facilitated if the free spectral range of the Vernier filter is wider than or equal to the gain bandwidth. To obtain single-frequency output with high side mode suppression the side peaks of the Vernier filter need to be sufficiently suppressed and the main filter peak should be narrowband. A main condition always remains, that the feedback of the $\mathrm{Al_{2}O_{3}}$ circuit at the main filter peak is high enough for reaching laser threshold with the available small-signal gain. Fulfilling all of these conditions requires a functional circuit design where all circuit parameters are properly set. Such conditions are the free spectral ranges of the micro-ring resonators used in the Vernier filter, the power coupling coefficients of the ring resonators ($\mathrm{\kappa}^2$), the mode matching between the amplifier and feedback chips, and the variable strength of the optical output coupling from the laser resonator. The amplifier's gain bandwidth determines the Vernier filter's free spectral range and the overall allowed feedback loss, while the amplifier's mode profile determines what minimum loss can be achieved in edge coupling and hybrid integration of the chips. Measuring the propagation loss of fabricated test structures determines the optimal coupling strength to the ring resonators that would maximize the peak filter transmission.\par
The selected amplifier is an InGaN/GaN superluminescent diode (SLED) with a center wavelength of 405 nm. Experimental data for the amplifiers are available from the manufacturer (Exalos AG).  To work as a double pass amplifier, the diode is highly-reflective coated (vs. air $>$95\%) on its back facet and anti-reflective coated (vs. air $<$0.1\%) on the facet facing the $\mathrm{Al_{2}O_{3}}$ chip. When operated without feedback and driven with a current of 78 mA the output is 10 mW of amplified spontaneous emission (ASE) with a 3-dB emission bandwidth of approximately 3.4 nm. The manufacturer specified mode field diameter is 1.87$\times$0.6 $\mathrm{\mu m}$.\par
For quantifying the propagation loss in the fabricated waveguides, we perform single-pass transmission measurements , using a set of waveguide spirals with weak curvature and different lengths (see Methods). The measurements yield a propagation loss of $\mathrm{\alpha}$ = 2.8 $\pm$ 0.3 dB/cm at a wavelength of 405 nm for the cross-section depicted in Fig. \ref{fig:1}b. The ASE emission bandwidth was used to set the radii of the Vernier resonators to (radius $\mathrm{R_1}$ = 150 $\mu$m and radius $\mathrm{R_2}$ = 153 $\mu$m), to obtain a Vernier spectral range of 4.84 nm (see Fig. \ref{fig:1}c). As a direct transmission measurement for fabricated Vernier filters is difficult, we measure instead the fabricated cross sections and tunnel gaps of the couplers using a scanning electron microscope (SEM) and use the retrieved geometry with index data to simulate the strength of couplers and Q-factors of ring resonators. The Vernier filter that was selected for most of the laser characterization experiments we obtain a power coupling of $\mathrm{\kappa^2}$ = 5.9\%. This Vernier filter would result in the narrowest spectral filtering, while still providing sufficient feedback to the amplifier to bring the laser above threshold. Using both parameters, $\alpha$ and $\kappa$, we conclude that both resonators are near-critically coupled with intrinsic and loaded Q-factors of 427.000 and 143.000, respectively, yielding a feedback filter resolution of approximately 3.3 GHz (full-width half maximum of Vernier resonance). The transmission function of this Vernier-filter is plotted in Fig. \ref{fig:1}c. It can be seen that the individual resonators with a free spectral range of about 97 pm open a Vernier free spectral range that is much wider, about 4.84 nm. This ensures that the filter passes only a single wavelength within the amplifier emission spectrum, qualitatively represented by the dashed outline of a measured ASE spectrum in Fig. \ref{fig:1}c. It can also be seen that the next-adjacent Vernier filter side mode (VFSM, see inset in \ref{fig:1}c) is well suppressed (3.3 dB below main peak), to maximize the side mode suppression during laser operation. For light passing resonantly through the Vernier filter, we calculate an optical roundtrip length of the laser cavity of 55.9 mm, which means the nearest cavity mode is at 5.36 GHz from the central lasing mode.\par
For optimum coupling between modes in the InGaN amplifier chip and $\mathrm{Al_{2}O_{3}}$ feedback chip, lateral inverse tapers are designed near the facets of the feedback chip. Limited by the elliptical shape of the amplifier output mode, the maximum theoretical coupling between both chip modes is calculated to be 91\%. Fresnel reflections from the interface of the chips into the amplifier are suppressed by letting the waveguides form a slight, index-matched angle with regard to the facet normal. For efficient coupling to output fibers, inverse tapers were designed at the output facet of the feedback chip. Other functionalities included in the design of the feedback chips are an adjustable phase section (PS), phase shifters for fine-tuning the optical length of the ring resonators ($\mathrm{R_{1}}$ and $\mathrm{R_{2}}$), an adjustable Mach-Zehnder interferometer (MZI) outcoupler (OC) formed by two 50\% directional couplers, and a phase section behind the outcoupler. Tuning of these circuit components is realized via the thermo-optic effect by placing thin-film resistive microheaters on the chip (as indicated in Fig. \ref{fig:1}a). Tuning the phase section allows for spectrally aligning a laser cavity mode with a resonance of the Vernier filter. Tuning of ring resonators ($\mathrm{R_{1}}$ and $\mathrm{R_{2}}$) selects a particular feedback wavelength. The main purpose of variable outcoupling is to enable maximum output power at each drive current of the laser diode and heaters of the Vernier filter. Varying the coupling is achieved by applying a current to one of the heaters at the arms of the MZI in the outcoupler. Characterization of test MZI structures showed that the outcoupling of the laser can be adjusted between approximately 10 and 90\%. To reduce ASE noise in the output, the output is taken from the filtered light from the Vernier loop mirror that is directed back to the laser amplifier. The phase section behind the outcoupler may be used to prevent injection locking of the laser frequency via unwanted reflections such as as from chip and fiber facets.
\subsection*{Hybrid integration and packaging}
Characterization of separate waveguide circuits can be carried out with manually aligned stages, such as in waveguide loss measurements using free-space fiber-to-chip coupling. Manual alignment of edge coupled feedback chips for spectral narrowing via self-injection locking can reveal ultra-low, Hertz-level intrinsic linewidths, however, vibration and drift of aligment stages typically limits stability to time scales shorter than milliseconds \cite{Jin2021a}. In contrast, long-term frequency and power stability, and inherent robustness to external perturbations, require mutual bonding of chips, hermetic sealing and temperature stabilization \cite{Ranno2022}. In infrared lasers, the laser cavity mode is allowed to propagate through bonding materials, and hermetic sealing is not essential. This is demonstrated with hybrid integration \cite{VanRees2020} and heterogeneous integration \cite{Tran2021a}. However, at the much higher UV photon energies, bonding materials need to be kept out of the laser mode and hermetic sealing is instrumental for long-term stable output.\par
\begin{figure}[!ht]
    \centering
    \makebox[\textwidth][c]{\includegraphics[width=180mm]{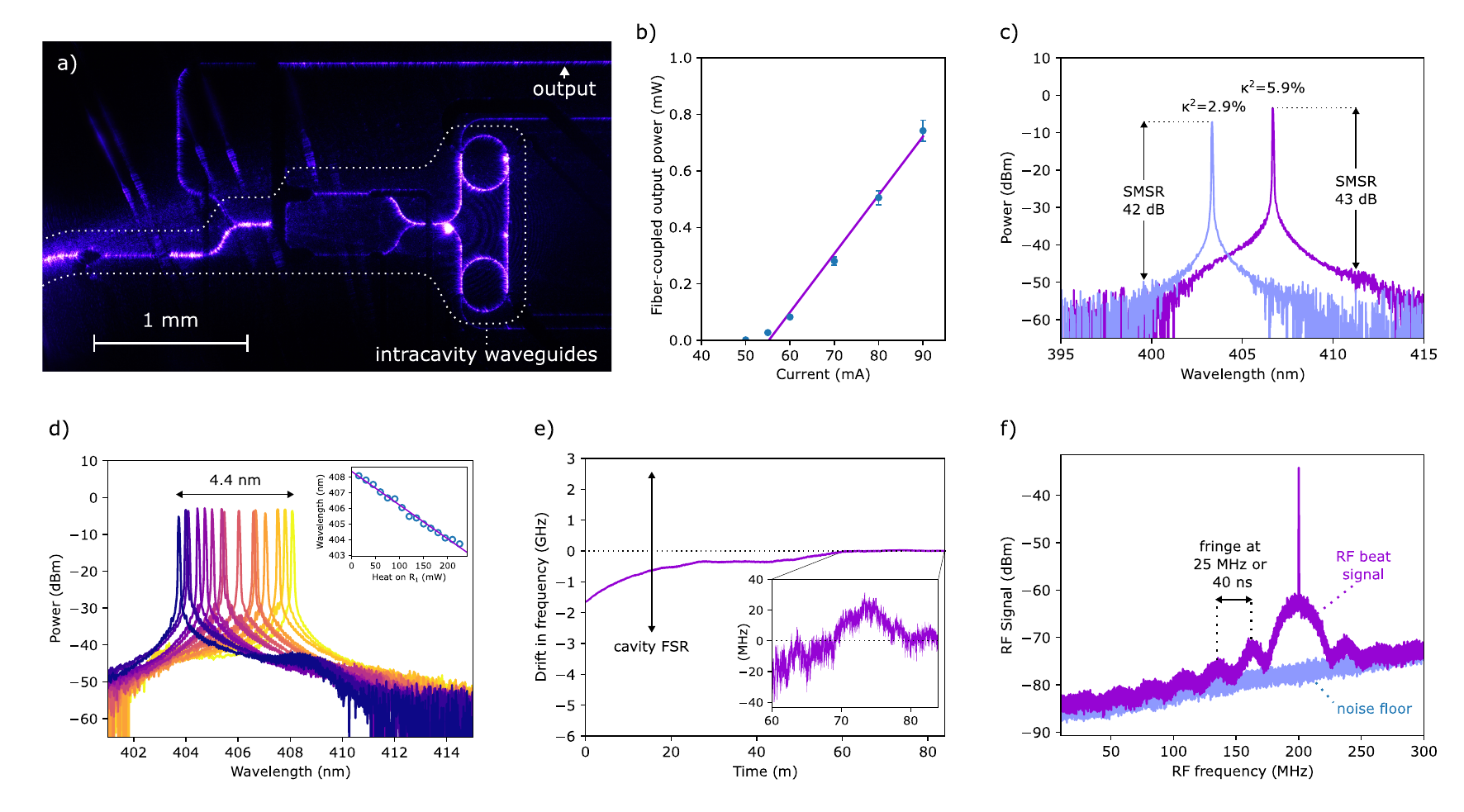}}%
    \caption{\textit{a) Microscopic picture of the $\mathrm{Al_{2}O_{3}}$-based hybrid laser in operation, showing scattered light from intracavity waveguides. Some parts of the waveguides are less bright due to the metal heater layers obscuring parts of the waveguide circuit. b) Fiber-coupled output power versus drive current, showing a laser threshold and slope efficiency of 55 mA and 20.7 $\pm$ 1 $\mu$W/mA, respectively. The fiber-coupled output power reaches a maximum of 0.74 $\pm$ 0.04 mW, correcting for fiber-to-chip coupling losses (6.7 $\pm$ 2.1 dB, see Methods) this corresponds to about 3.5 mW ($\pm$ 50\%) on-chip power in the output waveguide. c) Optical spectrum showing single wavelength operation with a sidemode suppression ratios (SMSR) of 43 dB ($\kappa^2$ = 5.9\%) and 42 dB ($\kappa^2$ = 2.9\%). d) Superimposed laser spectra shown for various heater settings of ring $R_{1}$, here, the fiber-coupled output power varies between 0.31 and 0.52 mW. The laser tunes in wavelength more than 4.4 nm from 408.1 nm to 403.7 nm, shown by center wavelength versus heating power plotted in the inset. Each tuning step is followed by automated optimization for the output power. The slope of the linear fit is -21.5 $\pm$ 0.5 pm/mW. e) High-resolution wavelength meter recording the laser wavelength over time. Recording begins right after start-up, when the laser is still far from thermal equilibrium. After an initial drift over 1.5 GHz, the frequency becomes constant with small residual fluctuations in the order of a few tens of MHz. No mode-hops are observed and the drift remains well below the cavity FSR. We attribute the high stability to hybrid integration of the chips and hermetic sealing. f) RF beat signal from a delayed self-heterodyne set-up for our laser, measured at 405 nm wavelength using a 7.9 m fiber delay (purple). The noise floor is recorded by blocking the photodiode from any input (blue). From the fringes in the RF beat signal we can extract a laser linewidth in the order of 25 MHz or lower.}}
    \label{fig:2}
\end{figure}

In total, three lasers were hybrid integrated and packaged with hermetic sealing in a standard 14-pin butterfly housing with electrical wirebonding and with feed-throughs for the output fibers and electrical signals (see Methods). The package contains a thermistor and Peltier element for temperature control of the laser. One packaged variant is sealed in a nitrogen atmosphere and equipped with a glass lid. This allows for visual inspection of the circuit in operation, such as seen in Fig. \ref{fig:1}e, \ref{fig:1}f and \ref{fig:2}a. For the other variants, improved hermetical sealing is realized by sealing the butterfly housing with a seamwelded metal lid which results in a significantly better hermetic sealing. On top of the improved sealing this variant uses an argon atmosphere, which might give an extended lifetime over a nitrogen buffer gas \cite{Schoedl2005a}. Feedback chips with two different coupling strengths between bus waveguides and and microring resonators were packaged ($\kappa^2$ = 5.9\% and 2.9\%).
\subsection*{Laser characterization}
 When the InGaN-$\mathrm{Al_{2}O_{3}}$ laser is succesfully brought in operation, the intracavity waveguides in the circuit light up brightly, see Fig. \ref{fig:1}f and \ref{fig:2}a. To record the fiber-coupled output power, for each measurement we make small readjustments of the heaters on the micro-resonators ($\mathrm{R_{1}}$, $\mathrm{R_{2}}$), the phase section (PS) and MZI-based outcoupler (OC). This optimization process is recursive and usually requires a run of one to three times for all parameters to find the maximum output power while maintaining single wavelength operation with high sidemode suppression. All heater adjustments can be carried out automatically via multichannel computer control of the heaters. The laser temperature control, a PID control loop using the thermistor and Peltier element in the package, is set to 20 \degree C for all measurements.\par
The fiber-coupled output power is shown in Fig. \ref{fig:2}b, we find a laser threshold current of 55.0 mA, above which the output increases linearly with pump current. The fiber-coupled, optical power reaches a value 0.74 $\pm$ 0.04 mW for the maximum drive current that we apply (90.0 mA). Correcting for losses at the chip-to-fiber coupling interface (estimated at 6.7 $\pm$ 2.1 dB, see Methods), the laser generates about 3.5 mW ($\pm$ 50\%) on-chip power in its output waveguide. At maximum output power the monitored wavelength is 405.5 nm. We note that the fiber-coupled, near UV output levels are up to two orders of magnitude higher than what has been reported recently with self-injection of stage-coupled Fabry-Perot lasers to silicon nitride feedback chips \cite{Corato-Zanarella2022}. Such large difference in output can have various reasons, however we expect that the main contributors to high output power are permanent bonding and much better mode matching between amplifier and feedback chip.\par
Wide-range near UV output spectra are recorded with an optical spectrum analyzer (OSA). Typical emission spectra as in Fig. \ref{fig:2}c display single-wavelength operation with high sidemode suppression ratios (SMSR) of about 42 dB  and 43 dB, which is two orders of magnitude more than previously reported around this wavelength range \cite{Siddharth2022,Corato-Zanarella2022}. These recordings indicate a full width at half maximum laser linewidth below the resolution limit of the optical spectrum analyzer (about 90 GHz). We perform coarse tuning of the laser by varying the heater power for one of the microring resonators ($\mathrm{R_{1}}$). Figure \ref{fig:2}d shows superimposed spectra of the laser at 14 coarse steps along its entire tuning range, with the inset showing the tunability of the center wavelength from 408.1 nm to 403.7 nm. The data shows a wavelength coverage of 4.4 nm, not exceeding the designed free spectral range of the Vernier (4.84 nm, Fig. \ref{fig:1}c).\par
To verify the promise of long term passive stability inherent to hybrid integration of lasers \cite{Epping2020}, we characterized the passive wavelength stability of the laser. Next to residual thermal drift, an important question is whether the laser would exhibit longitudinal mode hops. Such mode hops can impose limitations in electronic frequency locking to reference cavities or to absolute reference absorption lines. A mode hop changes the laser frequency by at least one free spectral range of the laser cavity,  which is about 5.36 GHz for this laser. Immunity from environmental perturbations can thus be judged by comparing the laser longer-term frequency drift with one cavity free spectral range. To measure the laser stability the wavelength was recorded over time using a high resolution wavelength meter. The result of such a measurement, at an average sampling time of 12 ms, is shown in Fig. \ref{fig:2}e. The figure shows that the laser was mode-hop-free for at least 84 minutes, which was also the entire duration of recording, and drifting less than 1.6 GHz. In the final 24 minutes the laser has reached thermal equilibrium with the environment, here a standard optical laboratory, and showed a drift of less than 30 MHz. Recording longer time traces does not appear problematic, however, the degradation times (inherent to all GaN-type diode lasers) are not known and presently not subject to systematic investigations.\par
In order to determine the laser linewidth with higher resolution than 90 GHz (from optical spectrum analyzer measurements), we use a delayed self-heterodyne experiment with a 7.9 meter long optical fiber as delay line (see Methods). A typical RF spectrum is displayed in Fig. \ref{fig:2}f (purple trace). The presence of fringes proves that the laser coherence time is approximately similar to or longer than delay time provided by the fiber. However, the fringe contrast in the line wings is strongly masked by the noise floor (blue trace). This prevents an evaluation for Lorentzian line shape components \cite{Richter1986} and thus requires that the signal-to-noise ratio is increased in further experiments. Nevertheless, the fringe period can be extracted reliably as 40 ns, which corresponds to the 7.9 m fiber delay. This gives a lower bound for the coherence time, i.e., the full width of the laser spectrum is in the order of 25 MHz or lower.
\subsection*{Summary}
In this work we have shown the design, fabrication and operation of the first near UV hybrid integrated diode laser using the deep-UV capable $\mathrm{Al_{2}O_{3}}$ material platform. The fabricated $\mathrm{Al_{2}O_{3}}$ chips show a propagation loss of 2.8 $\pm$ 0.3 dB/cm. For the first time a 405 nm InGaN amplifier was hybrid integrated with an $\mathrm{Al_{2}O_{3}}$ waveguide circuit, ensuring a UV-compatible bonding between both chips in a hermetically sealed environment. This approach maximizes the frequency stability and durability of the device. The laser shows a maximum output power of 0.74 mW, on-chip about 3.5 mW, and tunes more than 4.4 nm from 408.1 nm to 403.7 nm. The laser shows high passive frequency stability, operating without mode hops over more than an hour, settling to residual frequency deviations of a few ten MHz per 20 minutes. Delayed self-heterodyne measurement indicates that the full laser linewidth is in the order of the inverse delay time (25 MHz) or below, presently a low signal-to-noise ratio in heterodyne detection excludes a closer quantification. Continuing this work we aim on improved linewidth measurements via higher signal-to-noise detection. Our experiments show that the $\mathrm{Al_{2}O_{3}}$ platform newly engaged here for near UV, complex and tunable feedback circuits opens the path for chip integrated UV applications based on fully functional hybrid integrated diode lasers.

\clearpage
\section*{Methods}
\subsection*{Design and fabrication}
Simulation of optical modes (2D finite element methods, Lumerical), sidewall scattering integrals, and mode overlap integrals were performed to select the core cross-section as shown in Fig. \ref{fig:1}b. With this cross-section the expected propagation loss from scaling is $\mathrm{<}$4 dB/cm and calculations show tight bend radii down to 80 $\mu$m result in negligible bend radiation loss. The named cross-section was used to design directional couplers with nominal values $\mathrm{\kappa^2}$ of 1, 2, 5 and 10, 20 and 50\%. Manufacturers specifications on the the modefield diameter for the diode amplifier and UV fibers were consulted to design linear inverse tapers.\par
The $\mathrm{Al_{2}O_{3}}$ waveguides, fabricated by the Integrated Optical Systems group in the MESA\textsuperscript{+} cleanroom (University of Twente), starts by depositing a 110 nm $\mathrm{Al_{2}O_{3}}$ layer using an optimized RF reactive sputter deposition process \cite{VanEmmerik2020}, onto an 8 $\mathrm{\mu}$m thick thermally oxidized 10 cm diameter silicon wafer. A chemical mechanical polishing step is used to reduce the surface roughness of the deposited $\mathrm{Al_{2}O_{3}}$ layer, reducing the layer thickness to the targeted 100 nm. Next, the substrate is coated with negative e-beam resist (AR-N 7520). Using a Raith EBPG5150 e-beam lithography system the waveguide layer is written in the resist. After the e-beam write, the pattern is developed using AR-300-47 developer. The resulting patterns are etched into the $\mathrm{Al_{2}O_{3}}$ using an Oxford PlasmaPro 100 Cobra (reactive ion etching). Afterwards the resist is stripped by oxygen plasma using a TEPLA 300. The resulting waveguides are fully buried by an 8 $\mathrm{\mu}$m thick $\mathrm{SiO_{2}}$ cladding. To implement thermo-optic tuning at various locations on the feedback chips, resistive heaters are fabricated, by deposition and structuring of a 10/10 nm Cr/Pt layer topped with a 300 nm Au layer. All three layers are patterned using a lift-off process. To create the heaters, the gold layer is etched away leaving only the two thin and high electrically resistive Cr and Pt layers. For the $\mathrm{Al_{2}O_{3}}$ thin film the material index was measured using an ellipsometer (Woollam M-2000UI) in the wavelength range from 600 - 1600 nm. Cauchy's equation, $\mathrm{n(\lambda)=A + B/\lambda^2}$ with $\mathrm{\lambda}$ in $\mathrm{\mu}$m, was fitted to the ellipsometer data with A = 1.6848 $\pm$ 0.009 and B = 0.0119 $\pm$ 0.002. The function was extrapolated to obtain the material index in the near UV spectral range.
\subsection*{Characterization of $\mathbf{Al_{2}O_{3}}$ feedback chip}
\begin{figure}[t]
    \centering
    \makebox[\textwidth][c]{\includegraphics[width=50.5mm]{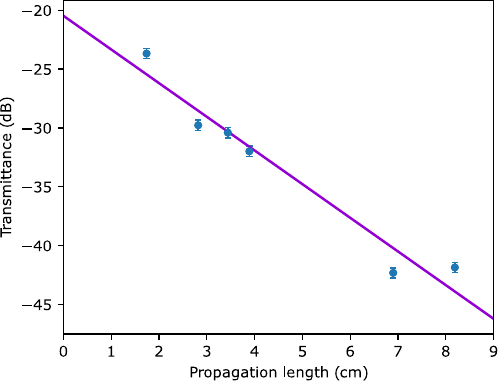}}%
    \caption{\textit{Transmittance ($P_{out}/P_{in}$) of 405 nm light through various paths and spirals of different propagation length. The linear slope indicates a propagation loss of 2.8 $\pm$ 0.3 dB/cm and from the y-axis crossing a 10.2 $\pm$ 0.8 dB fiber-to-chip coupling loss is obtained.}}
    \label{fig:3}
\end{figure}

Several chips were fabricated with test structures to characterize individual building blocks of the waveguide feedback circuit. First, to characterize the straight propagation loss, several spirals and straight waveguides were investigated. These structures used a minimum bend radius of 150 $\mathrm{\mu m}$, which ensures negligible bend radiation loss while maintaining a small footprint circuit. Using a 405 nm Fabry-Perot diode laser (QPhotonics) light was fiber coupled (PM-S405-XP) into the chip, passing through a spiral or path (Fig. \ref{fig:1}d). At the output side of the chip the transmitted power was measured (Thorlabs, S150C photodiode). The transmittance ($\mathrm{P_{out}/P_{in}}$) as a function of the on-chip propagation length is shown in Fig. \ref{fig:3} and yields an average propagation loss of 2.8 $\pm$ 0.3 dB/cm and a fiber-to-chip coupling loss of 10.2 $\pm$ 0.8 dB/facet. To estimate the loss of the inverse tapers, test structures were investigated where light propagates sequentially through a number of adiabatic tapers lined up as a waveguide and showed a loss of the inverse tapers below the measurement accuracy. To determine the phase tuning of light using the fabricated heaters, another test structure with an MZI and a thermal heater on each arm revealed that about 350 mW of electrical power is needed for a phase shift of 2$\mathrm{\pi}$ in the guided light. Scanning electron microscope (SEM) images allowed verification of the cross-section, gaps at directional couplers and tapers on various parts of the photonic chip. These cross-sectional images also verified that the waveguide sidewall is at the designed right angle with the base (approximately 90$\degree$).
\subsection*{Hybrid integration and packaging}
The hybrid integration and packaging process was carried out in conjunction with PHIX B.V. and with support from Lionix International B.V. First the individual $\mathrm{Al_{2}O_{3}}$ chips were diced from the wafer and polished. The right side of the feedback chip (Fig. \ref{fig:1}d) was polished at an 8$\degree$ vertical angle to match the angle of the fiber array, minimizing any back reflections into the laser mode from this interface. Also here, the waveguide width is tapered down to maximize the mode matching with the large and circular 3.3 $\mu$m diameter mode of the PM-S405-XP fiber, showing a theoretical coupling efficiency of 59\% limited by the elliptical modeshape in the waveguide. Here, a fiber array with five of these fibers was used. For the active temperature control a 10 k$\ohm$ thermistor was added to the SLED submount.\par
After preparing both chips, the SLED is aligned and butt-coupled to the $\mathrm{Al_{2}O_{3}}$ feedback chip. When optimum alignment is reached, the chips are hybrid integrated by bonding using a UV curable epoxy. The epoxy is applied such that the near UV optical mode remains free from epoxy. On the output side of the feedback chip the fiber array is aligned for maximum coupling and bonded using an epoxy, again ensuring a free optical path between the feedback chip and fibers. Subsequently a Peltier element is mounted to the bottom of a shared substrate, to be used with the thermistor for the thermal control of the laser. Afterwards the full assembly is placed in a standard 14-pin butterfly package. The electrical connections for the cathode and anode of the SLED and heaters on the $\mathrm{Al_{2}O_{3}}$ feedback chip are wirebonded to the pins of the butterfly package. The Peltier connectors are soldered to the remaining two pins of the package.\par
The final step is hermetic sealing of the butterfly package. For each package a getter material is attached to the inside of the lid before hermetically sealing the package. The getter material absorbs any residual gas (from outgassing of the epoxy, like volatile organic compounds and moisture after sealing). The laser in the pictures (Fig. \ref{fig:1}e, \ref{fig:1}f, \ref{fig:2}a) is a package with a glass lid in a nitrogen environment ($\mathrm{\kappa^2}$ = 5.9\%). The measurements shown in this work are carried out with a seamwelded, argon atmosphere laser ($\mathrm{\kappa^2}$ = 5.9\%), except the measurement for one of the optical spectra shown in Fig. \ref{fig:2}c (seamwelded, argon atmosphere, $\mathrm{\kappa^2}$ = 2.9\%).
\subsection*{Experimental methods}
Waveguide heaters are driven with a high-precision, low-noise, multichannel power supply (Chilas B.V., Tunable Laser Controller, TLC). The TLC also contains a PID control loop which controls the temperature of the laser with the thermistor and Peltier element in the butterfly housing. The TLC is equipped with a USB interface to receive serial commands from a PC. Together with in-house software, photodiodes and other lab equipment, optimization and control of the laser can be fully automated. Depending on availability several current sources were used: Toptica DLC Pro, ILX Lightwave LDX-3620a and Thorlabs LDC205B.\par
Output powers are measured by connecting the output fiber to a calibrated photodiode (Thorlabs, S150C photodiode). The calibration accuracy of the photodiode is 5\% in this spectral range. A 405-nm 90/10 custom fiber splitter (Thorlabs with S405-XP fiber) is used for simultaneous monitoring of spectral properties and output power. To find the conversion for the fiber-coupled output power to the on-chip power, we first note the measured fiber-to-chip coupling loss, which is 10.2 $\pm$ 0.8 dB (Fig. \ref{fig:3}). This value is measured with an unpolished chip facet, whereas the integrated laser facets are finely polished which should reduce the fiber-to-chip coupling loss by an estimated 3.5 $\pm$ 1.1 dB \cite{Jette-Charbonneau2008}. Finally, the 5\% ($\sim$ 0.2 dB) measurement error for the photodiode can be included in this conversion error. These estimates bring the total conversion factor from fiber-coupled output to on-chip power to +6.7 $\pm$ 2.1 dB. Therefore, we find an on-chip power of about 3.5 mW ($\pm$ 50\%).\par
The measurements for the spectra, wavelength stability and spectral linewidth are all recorded at the maximum current of 90 mA. Laser output spectra are recorded with an optical spectrum analyzer (Ando AQ6315A, approximately 50 pm or 90 GHz resolution at 405 nm). To measure the optical spectra the 90/10 fiber splitter is used to simultaneously monitor the output power of the laser with a photodiode, which enables to use our optimization software for controlling the laser. The nominally 90\% arm of the fiber splitter is connected to the spectrum analyzer and the spectra are corrected (due to the splitting and loss in fiber splitter) in post-processing for the photodiode measured, fiber-coupled output power of the laser. Long-term measurements of passive laser frequency stability rely on a high-resolution wavelength meter (HighFinesse WS-U 1645, wavelength deviation sensitivity of 0.5 MHz ).\par
Delayed self-heterodyne measurements, using existing methods  \cite{Richter1986}, are carried out with a fiber delay length of 7.9 m (S405-XP fiber), an acousto-optic modulator (at 200 MHz, G$\mathrm{\&}$H FiberQ) and a balanced photodiode (Thorlabs PD435A-AC), with signals being recorded with an electrical spectrum analyzer (Keysight CXA N9000B). To match the polarization of the signal from both arms of the set-up, a fiber polarization controller is used (Thorlabs FPC560 with an S405-XP fiber). The recorded RF spectrum of the beatnote is shown as the purple trace in Fig. \ref{fig:2}f.  The measurement time of an RF frequency sweep is 67 ms and the raw data is averaged 20 times. The fringe period is extracted from the RF signal as 40 ns, which corresponds well with a 7.9 m fiber delay length. The noise floor is recorded when the light to the sensor is blocked (blue trace, Fig. \ref{fig:2}f).

\clearpage
\printbibliography


\end{document}